\documentclass[twocolumn]{autart}

\usepackage{amsmath} 
\usepackage{mathrsfs}
\usepackage{amssymb}
\usepackage{graphicx}
\usepackage{caption}
\usepackage{subcaption}
\usepackage{parskip} 
\usepackage{setspace} 
\usepackage{emptypage} 
\usepackage{multicol} 
\usepackage{booktabs}
\usepackage{multirow}
\usepackage{enumitem}   
\usepackage{xcolor}
\usepackage{mathtools}
\DeclareMathOperator{\vect}{vec}
\newtheorem{rmk}{Remark}

\newcommand{\ie}{{\em i.e.}}

\makeatletter
\def\ps@copyright{%
  \let\@mkboth\@gobbletwo
  \def\@oddhead{}%
  \let\@evenhead\@oddhead
  \def\@oddfoot{}%
  \let\@evenfoot\@oddfoot
}
\makeatother

\begin{document}

\begin{frontmatter}

\title{Continuous-time Predictor-Based Subspace Identification with Hermite basis expansions} 

\thanks[footnoteinfo]{
}

\author[PoliMI]{Jose Antonio Rebollo}\ead{jrebollo@us.es},    
\author[PoliMI]{Enrico Barbiero}\ead{enrico.barbiero@polimi.it},                
\author[PoliMI]{Marco Lovera}\ead{marco.lovera@polimi.it}  
                                           
\address[PoliMI]{Department of Aerospace Science and Technology,
	Politecnico di Milano, Via La Masa 34, 20156 Milano, Italy}


\begin{abstract}                          
In this paper the problem of continuous-time subspace identification for Linear Time Invariant (LTI) systems is considered and a method which directly identifies a continuous-time  state-space form is proposed. First, Hermite basis functions are used to project signals and obtain a finite number of Hermite coefficients. By exploiting recursive relations and time derivative properties of the Hermite basis functions, an expression of the derivative operator is obtained. The latter is then recursively applied, ensuring that the state-space matrices remain in continuous-time form and making the system suitable for the implementation of steps which are akin to those of the Predictor-Based Subspace IDentification (PBSID) method.
This new method, hereby called the Hermite-Domain PBSID (HD-PBSID) method, has the further advantage of avoiding time-shifts by properly scaling the input and output signals. 
The performance of the proposed approach is illustrated in a simulation study aimed  at showing the accuracy of the estimates and at  comparing the HD-PBSID method and the Laguerre-projections based Continuous-Time PBSID (CT-PBSID) algorithm.     

\end{abstract}

\end{frontmatter}

\section{Introduction}
System identification seeks to build accurate
mathematical models of system dynamics from experimental data.
This field is of great interest for countless applications for which reliable knowledge of plant dynamics is critical for safe and effective operation, control design, and performance analysis.
Frequency-domain methods characterize a system by how it responds across frequency, typically by estimating a Frequency Response Function (FRF) from measured input–output data \cite{ljung1999system}. In practice, this is often done by exciting the system with periodic or broadband signals (e.g., sine sweeps, multisine signals, or random noise), computing spectral estimates such as auto- and cross-power spectral densities, and forming nonparametric FRF estimates. These approaches are especially attractive when the goal is to understand resonance behavior, bandwidth, and disturbance attenuation.
Time-domain methods fit parametric models directly to measured input–output time series. While transient excitations such as steps, pulse trains, and impulse-like signals are common, time-domain identification can also use broadband periodic inputs (e.g., multisine) or even sweep signals, provided they sufficiently excite the system dynamics. These methods often aim to identify state-space representations, which are convenient for MIMO modeling and can be used directly for control design. Subspace identification is a prominent family within time-domain methods: it estimates MIMO state-space models from input–output data mainly through numerical linear algebra (projections and singular value decomposition), typically avoiding the nonlinear iterative optimization required by prediction-error approaches \cite{van1996subspace,verhaegen2007filtering}.
In particular, the survey by Van der Veen, van Wingerden, Lovera, and Verhaegen \cite{van2013closed} provides a comprehensive overview of several subspace identification methods that were proposed to tackle the closed loop identification challenge. For instance, the Multivariable Output-Error State-space (MOESP) method \cite{VerhaegenMOESP} estimates system dynamics by projecting the input-output data onto the orthogonal complement of the input and innovations subspaces. This projection is achieved through an $RQ$ factorization of the input-output block Hankel matrices, followed by a singular value decomposition (SVD) to determine the system order and extract the state-space matrices. The N4SID (Numerical Subspace State-Space Identification) method \cite{Overschee1996SubspaceIF} uses a different approach, directly projecting the input-output data onto subspaces associated with non-steady-state Kalman filter banks to compute the state sequences. These state sequences, representing optimal predictions of the system states, are then used to construct the system matrices. One of the most widely used subspace identification techniques in closed-loop applications is the Predictor Based Subspace IDentification (PBSID) algorithm and its variants \cite{jansson2003subspace,chiuso2007relation}. This method, in contrast, combines subspace projections with prediction error minimization. It identifies the system by first estimating prediction errors using oblique projections of the data and then solving a least-squares problem to determine the state-space matrices. This hybrid approach improves model accuracy, especially in closed-loop settings, by explicitly accounting for the feedback structure during the estimation process. Other approaches include using specific excitations or Instrumental Variables (IV) \cite{chou1997subspace,chou2004direct}, also detailed in \cite{van2013closed}, or applying methods that require controller information.

While the vast majority of identification methods focus on discrete-time systems, there is a need for directly identifying continuous-time models, especially when dealing with stiff systems or non-uniformly sampled data \cite{garnier2008identification,garnier2011editorial}. 
Converting discrete-time models to continuous time can be problematic because discretization may yield non-unique continuous-time realizations, i.e., the mapping from continuous to discrete time is not one-to-one \cite{farahBOUNDED_ERROR_CTMODELS}.

Several approaches for continuous-time identification have been proposed, as detailed in \cite{bergamasco2012continuous}. These often rely on basis functions to avoid the numerical issues related to high order derivative computation.
This includes transformations based on  generalized orthonormal bases, as proposed by \cite{ohta2004continuous}, \cite{ohta2005realization}. This family of functions includes Kautz-type bases and the Laguerre basis. The latter is also used in \cite{HAVERKAMP_MOESP} to develop a continuous-time version of MOESP, and in \cite{bergamasco2011continuous2} for the Continuous-Time PBSID (CT-PBSID). These algorithms have the common property of identifying continuous-time models without incurring nonbijective transformations. Two different strategies for the continuous-time to discrete-coefficient transformation for the CT-PBSID algorithm are available: one relies on convolutions with Laguerre filters, and the other on projections into a basis of Laguerre functions. The latter has been shown to provide better results in terms of bias \cite{Bergamasco2011}.      
One of the main characteristics of the Laguerre-projections CT-PBSID is the need for repetition of the projections by time-shifting the input and output signals. This is posed by the necessity of using a small set of Laguerre basis functions. Indeed, these functions tend to become numerically unstable when their order is increased, since the sampling time used in the projections is fixed and depends on the experimental signal data. The maximum-order Laguerre basis function is thus limited in settling time and hence requires time-shifting the signals.
However, the need for multiple shifts causes the algorithm to slow down when long experiments are considered. In addition, estimates of the asymptotic covariance of systems identified via CT-PBSID could be negatively affected by correlations between projected signals \cite{Bootstrap-based}. 
Another issue affecting the CT-PBSID algorithm is that it first identifies the state-space matrices in the Laguerre domain and then back-transforms them to the continuous-time domain, which may lead to additional computational errors.
Furthermore, the Laguerre basis functions depend on the pole of the all-pass transfer function that originates the Laguerre basis. The choice of the Laguerre pole significantly affects the accuracy of the models identified via CT-PBSID, since small variations in its value lead to large differences in the identified models. To address the large sensitivity to the selection of the pole, \cite{Verhaegen2017} proposes an identification method which uses Takenaka-Malmquist bases, of which Laguerre functions are a special subclass. 
However, this approach still depends on user-selected dynamic parameters related to these basis functions.

Hermite orthogonal basis functions provide a promising alternative because they do not rely on such parameters.
In addition, they satisfy recursive relations that allow basis functions of increasing order to be generated efficiently, while their time derivatives can be obtained directly from the same relations \cite{szego1959orthogonal}. 
When the input and output signals of a state-space system are projected onto Hermite basis functions, these properties enable an explicit representation of the derivative operator and its recursive application.
As a result, unlike Laguerre-projection CT-PBSID, the resulting method identifies the state-space matrices directly in continuous-time form. Furthermore, no time shifts are needed for the identification to be effective. We refer to this new method as the Hermite-Domain PBSID (HD-PBSID).

\section{Problem statement}
Consider the LTI stochastic CT system:
\begin{equation}
\label{eq: LTI stochastic CT system}
\begin{aligned}
dx(t) &= A x(t)\,dt + B u(t)\,dt + dw(t), \qquad x(0)=x_0,\\
dz(t) &= C x(t)\,dt + D u(t)\,dt + dv(t),\\
y(t)\,dt &= dz(t).
\end{aligned}
\end{equation}
where state $x \in \mathbb{R}^{n_x}$, input $u \in \mathbb{R}^{n_u}$ and output $y \in \mathbb{R}^{n_y}$, while $w \in \mathbb{R}^{n_x}$ and $v \in \mathbb{R}^{n_y}$ are, respectively, process and measurement noises, modeled as Wiener processes with the following incremental covariance:
\begin{equation}
\mathbb{E}\left\{
\begin{bmatrix}
dw(t)\\
dv(t)
\end{bmatrix}
\begin{bmatrix}
dw(t)\\
dv(t)
\end{bmatrix}^{T}
\right\}
=
\begin{bmatrix}
Q & S\\
S^{T} & R
\end{bmatrix}
dt
\end{equation}
The system matrices  $A \in \mathbb{R}^{n_x \times n_x}$, $B \in \mathbb{R}^{n_x \times n_u}$, $C \in \mathbb{R}^{n_y \times n_x}$, and $D \in \mathbb{R}^{n_y \times n_u}$ are such that $(A,C)$ is observable and $(A,[B,Q^{1/2}])$ controllable.
Sampled input/output data $\mathcal{S} \vcentcolon = \{u(t^u_1), u(t^u_2), \hdots, u(t^u_{N_u}),$ $ y(t^y_1), y(t^y_2), \hdots, y(t^y_{N_y})\}$ from system (\ref{eq: LTI stochastic CT system}) is available. The sequence of sampling instants can be non-equidistant and the input sequence may not coincide with the output sequence; however, it is assumed that $t^u_1 \equiv t^y_1$ and $t^u_{N_u} \equiv t^y_{N_y}$.
\\Under the above-mentioned assumptions, system (\ref{eq: LTI stochastic CT system}) can be rewritten in innovation form:
\begin{equation}
\label{eq:LTI stochastic system in innovation form}
\begin{split}
d\hat{x}(t) &= A\hat{x}(t)\,dt + Bu(t)\,dt + K\,d\xi(t), 
\qquad x(0) = x_0, \\
d\hat{z}(t) &= C\hat{x}(t)\,dt + Du(t)\,dt, \\
y(t)\,dt &= d\hat{z}(t) + d\xi(t),
\end{split}
\end{equation}
where the innovation $\xi(t) \in \mathbb{R}^{n_y}$ is a Wiener process, and $K$ is the Kalman gain.
\\The classical Laguerre-projections CT-PBSID method exploits Laguerre basis functions for discretizing system (\ref{eq:LTI stochastic system in innovation form}), while having the following byproducts:
\begin{enumerate}[label=(\roman*)]
        \item Signals $u$ and $y$ are repeatedly time-shifted and projected onto a finite set of Laguerre basis functions. This repeated shifting may induce correlations within the innovation sequence in the Laguerre domain.   
        \item The state-space matrices are first identified in the Laguerre domain. By calling them $A_o \; , \; B_o \; , \; C_o \; ,$ $ \; K_o \; , \; D_o$, the continuous-time matrices are then obtained with a transformation $\psi$ such that: 
        \begin{equation}
        \vect(\hat{A},\hat{B},\hat{C},\hat{K},\hat{D}) = \psi(\vect(A_o,B_o,C_o,K_o,D_o)),
        \end{equation}
        where the matrices are identified up to a similarity transformation $T \in \mathbb{R}^{{n_x} \times {n_x}} : T^{-1}AT , \; T^{-1}B , $ $\; T^{-1}K, \; CT , \; D$.
\end{enumerate}
The problem is to derive, using orthonormal Hermite basis functions and their differentiation recursions, a discrete-coefficient representation equivalent to (\ref{eq:LTI stochastic system in innovation form}) such that:
\begin{enumerate}[label=(\roman*)]
        \item The projections of the signals are obtained without the need of multiple time-shifts.
        \item The matrices $A,B,C,D,K$ remain in continuous-time form, and thus no transformation $\psi$ is necessary.
\end{enumerate}
Furthermore, the PBSID procedure must be reformulated for the resulting Hermite-domain representation.

In the following section the main properties of Hermite basis expansions is summarized and the equivalent system with Hermite-projected signals is derived, while the HD-PBSID method is presented and discussed in Section \ref{sec:hdpbsid}.

\section{Continuous-time to discrete-coefficient transformation via Hermite basis expansions}

\subsection{LTI systems and Hilbert spaces}
Consider the CT LTI system of equation (\ref{eq:LTI stochastic system in innovation form}). One can rewrite it in the following form:
\begin{equation}
\label{eq: CT LTI dynamics in innovation form with explicit derivative}
\begin{split}
\frac{d}{dt}\hat{x}(t) = A \hat{x}(t) + B u(t) + Ke(t) \\
y(t) = C \hat{x}(t) + D u(t) + e(t)
\end{split} \quad ,
\end{equation}
which is a formal signal representation that is suitable for the subsequent basis-expansion derivation.
\\Signals $\hat{x}(t), u(t), y(t), v(t), w(t) $ can be interpreted as elements of a continuous Hilbert space $\mathcal{L}^2(\mathbb{R})$ or $\mathcal{L}^2(\mathbb{R}^+)$. The state-space matrices are understood as operators between Hilbert spaces, here generally denoted as $ \Phi : \mathcal{L}^2 \rightarrow \mathcal{L}^2 $. The time derivative $d(\cdot)/dt$ is interpreted as a densely defined linear operator on a suitable Hilbert space; for instance,
$d(\cdot)/dt : \mathscr{D}(\mathbb{R}) \subset \mathcal{L}^2(\mathbb{R}) \to \mathcal{L}^2(\mathbb{R})$, where e.g. $\mathscr{D}(\mathbb{R}) = H^1(\mathbb{R})$ is the first Sobolev space, that is the set of functions that are square-integrable and whose first derivative (in the weak sense) is also square-integrable.
\\The first main objective of this paper is to find a transformation that converts the CT LTI system (\ref{eq: CT LTI dynamics in innovation form with explicit derivative}) into an equivalent Discrete-Coefficient LTI (DC LTI) system.
\\In the discrete framework, signals $\tilde{x}, \tilde{u}, \tilde{y}, \tilde{v}, \tilde{w}$ are interpreted as elements of a discrete Hilbert space, namely $\ell^2(\mathbb{Z})$ or $\ell^2(\mathbb{N})$. The state-space matrices and the derivative are linear operators over the Hilbert space, generally denoted by $\tilde{\Phi} : \ell^2 \rightarrow \ell^2$.

\subsection{Schauder bases of $\mathcal{L}^2$}
\label{subsec: Schauder bases}
The objective of transforming (\ref{eq: CT LTI dynamics in innovation form with explicit derivative}) into an equivalent DC LTI system (\ref{eq: CT LTI dynamics in innovation form with explicit derivative}) can be achieved by expanding the signals using Schauder bases. 
\\A Schauder basis, or countable basis, for a Hilbert space $\mathcal{H}$ is a sequence $\{b_i\}_{i \in \mathbb{N}}$ of elements $b_i \in \mathcal{H}$ that satisfies
\begin{equation}
    f = \sum_{i \in \mathbb{N}} \tilde{f}_i b_i, \quad \forall f \in \mathcal{H}, \label{SeriesExpansion}
\end{equation}
and for which the coefficients $\tilde{f}_i$ are unique. A Hilbert space that admits a Schauder basis is called separable. Thus, a countable basis can be defined in a Hilbert space if and only if the space is separable. Both $\mathcal{L}^2(\mathbb{R})$ and $\mathcal{L}^2(\mathbb{R}^+)$ are separable spaces. Furthermore, any separable infinite-dimensional Hilbert space is isometrically isomorphic to $\ell^2(\mathbb{N})$. Let two spaces $\mathcal{H}_1$ and $\mathcal{H}_2$ be isometrically isomorphic, where for instance $\mathcal{H}_1$ may be $\mathcal{L}^2(\mathbb{R})$ and $\mathcal{H}_2$ may be $\ell^2(\mathbb{N})$. Thus, there exists a transformation $T : \mathcal{H}_1 \rightarrow \mathcal{H}_2$ which verifies the following:
\begin{enumerate}[label=(\roman*)]
\item Isometry: $T$ preserves the norm of vectors, namely, $\|f\|_{\mathcal{H}_1} = \|T(f)\|_{\mathcal{H}_2}, \forall f \in \mathcal{H}_1$. For Hilbert spaces, this implies that $T$ is unitary.
\item Isomorphism: $T$ is bijective and the algebraic structure of the spaces is preserved. For Hilbert spaces, this implies that $T$ is linear invertible.
\end{enumerate}
It follows that, by exploiting Schauder bases that have the property of being orthonormal, it is possible to provide a straightforward transformation $T$, based on projections, that associates signals in $\ell^2$ to signals in $\mathcal{L}^2$. In particular, the projection of a signal $f$ into the $i$-th orthonormal Schauder basis function is given by:   
\begin{equation}
\label{eq: projection on a Schauder basis}
    \tilde{f}_i = \langle b_i,f \rangle_{\mathcal{L}^2} = \int_{\mathbb{T}} f(t) b_i(t) \, dt,
\end{equation}
with $\mathbb{T} \subset \mathbb{R}$.
\\Classical orthonormal countable bases of $\mathcal{L}^2(\mathbb{R})$ and $\mathcal{L}^2(\mathbb{R}^+)$ are the Hermite functions and the Laguerre functions, respectively. Laguerre functions have been extensively studied for CT to DC transformations (see \cite{HAVERKAMP_MOESP}, \cite{bergamasco2011continuous2}), although their application is not derived from the theory of Hilbert spaces and Schauder bases but by using the methods of system lifting and Laguerre filtering. Instead, to the authors' knowledge, Hermite functions have not so far been explicitly considered for the CT system identification problem. The definition of the Hermite functions and of their useful properties will be provided in Subsection \ref{subsec: Hermite functions}.

\subsection{Construction of CT-DT transformations}
\label{Subsec: Construction of CT-DT transformations}
As pointed out in Subsection \ref{subsec: Schauder bases}, orthonormal Schauder bases $\{b_i\}_{i \in \mathbb{N}}$ enable associating $\ell^2$ signals to $\mathcal{L}^2$ signals.  
In addition, given an operator $\Phi : \mathscr{D} \subset \mathcal{L}^2 \to \mathcal{L}^2$, it is possible to discretize it using an orthonormal Schauder basis. For instance, consider the vectors $f,f'\in\mathcal{L}^2$ satisfying
\begin{equation}
     f' = \Phi f,
\end{equation}
then by expanding the vectors into their series, as in definition (\ref{eq: projection on a Schauder basis}), it follows that
\begin{equation}
    \sum_{i \in \mathbb{N}} \langle b_i, f' \rangle b_i = \Phi \sum_{i \in \mathbb{N}} \langle b_i, f \rangle b_i = \sum_{i \in \mathbb{N}} \langle b_i, f \rangle \Phi b_i.
\end{equation}
By projecting both sides on the basis element $b_j$, it follows that
\begin{equation}
    \langle b_j, f' \rangle = \sum_{i \in \mathbb{N}} \langle b_i, f \rangle \langle b_j, \Phi b_i \rangle.
\end{equation}
Consequently,
\begin{equation}
    \tilde{f}'_j = \sum_{i \in \mathbb{N}} \tilde{\Phi}_{ji} \tilde{f}_i,
\end{equation}
where $\tilde{f}'_j = \langle b_j, f' \rangle$ and $\tilde{\Phi}_{ji} = \langle b_j, \Phi b_i \rangle$. Therefore, this procedure transforms an operator in $\mathscr{D} \subset\mathcal{L}^2$ into an operator in $\ell^2$ in the coordinates of the given basis.  Note that as per the definition of Schauder bases, any transformed vector $\tilde{f} = \{\tilde{f}_i\}_{i \in \mathbb{N}}$ and operator $\tilde{\Phi}_{ji}$ are unique, that is, there is only one representation of a vector $f$ and operator $\Phi$ as a linear combination of the elements of a basis. 
By defining the vector of coefficients $\tilde{f'} = \{\tilde{f'}_j\}_{j \in \mathbb{N}}$, one can concisely write the relation $\tilde{f'} = \tilde{\Phi}\tilde{f}$, where $\tilde{\Phi}$ is the matrix containing all the terms $\tilde{\Phi}_{ji}$: 
\begin{equation}
\label{eq: transformed general operator}
\tilde{\Phi} = \begin{bmatrix}
\langle b_1, \Phi b_1 \rangle & \langle b_1, \Phi b_2 \rangle & \cdots \\
\langle b_2, \Phi b_1 \rangle & \langle b_2, \Phi b_2 \rangle & \cdots \\
\vdots & \ddots & \ddots
\end{bmatrix}.
\end{equation}
The general term $\tilde{\Phi}_{ji}$ is nonzero only if $\Phi b_i$ has some component in the direction of $b_j$. A nonzero $\tilde{\Phi}_{ji}$ term is said to `connect' the states $i$ and $j$. 
\\The uniqueness of the transformed vectors and operator, the possibility of calculating the matrix form of the transformed operator, and the choice of orthonormal Hermite basis functions as Schauder basis allow one to transform the derivative operator of system (\ref{eq: CT LTI dynamics in innovation form with explicit derivative}) from acting in $\mathscr{D} \subset \mathcal{L}^2 \to \mathcal{L}^2$ to acting in $\ell^2 \to \ell^2$, as shown in Subsection \ref{Differentiation operator}.

\subsection{Hermite functions}
\label{subsec: Hermite functions}
\begin{figure}[htbp]
    \centering
    \includegraphics[width=\columnwidth]{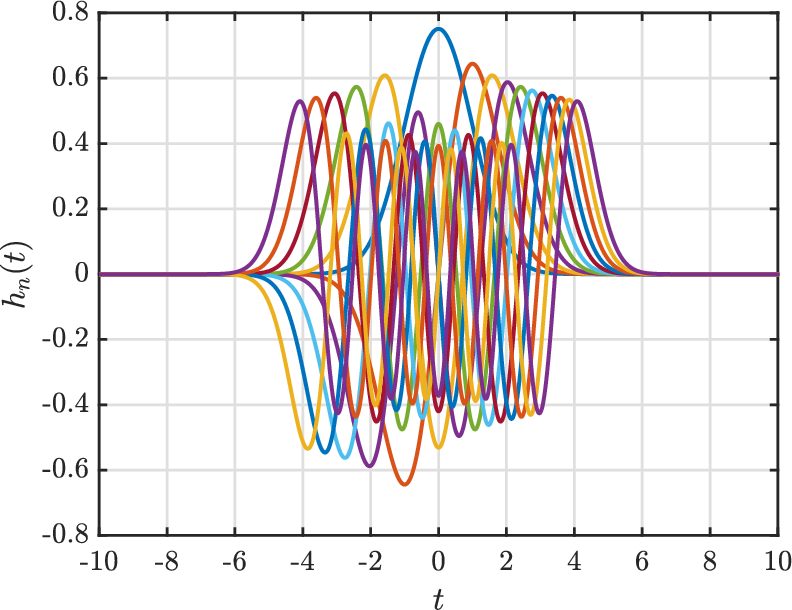}
    \caption{Hermite basis up to order 10.}
    \label{fig: Hermite functions up to order 10.}
\end{figure}
The normalized Hermite functions, $h_n(t)$, are defined using the Hermite polynomials\footnote{In the literature, the polynomials used in this paper are sometimes referred to as the \textit{physicist's Hermite polynomials}. An explicit expression for the $n$-th Hermite polynomial is the particularized Rodrigues' formula, $\mathrm{H}_n(t) = (-1)^n e^{t^2} \frac{d^n}{dt^n} e^{-t^2}$, although they can be generated using recurrence relations, as shown in this Section. } $\mathrm{H}_n(t)$ as
\begin{equation}
    h_n(t) = e^{-t^2/2} \frac{\mathrm{H}_n(t)}{\sqrt{2^n \sqrt{\pi} n!}}.
\end{equation}
The sequence $\{h_n(t)\}_{n \in \mathbb{N}}$ is an orthonormal Schauder basis of $\mathcal{L}^2(\mathbb{R})$, that is,
\begin{equation}
\langle h_n, h_m \rangle = \int_{-\infty}^{\infty}{h_n(t) h_m(t) dt} = \delta_{nm}, \forall n,m \in \mathbb{N}.
\end{equation}
Since Hermite polynomials $\mathrm{H}_n(t)$ satisfy the following recurrence relation:
\begin{equation}
\mathrm{H}_{n+1}(t) = 2x \mathrm{H}_n(t) - 2n \mathrm{H}_{n-1} (t),
\end{equation}
the recurrence relation for the Hermite functions turns out to be:
\begin{equation}
\label{eq: recurrence relations for Hermite functions}
h_{n+1} (t) = t \sqrt{\frac{2}{n+1}} h_n(t) - \sqrt{\frac{n}{n+1}} h_{n-1}(t).
\end{equation}
Additionally, the first two Hermite functions are given by:
\begin{align}
h_0 (t) & = \frac{\exp{\left(\frac{-t^2}{2}\right)}}{\pi^{1/4}}, \\
h_1 (t) & = \sqrt{2} t \frac{\exp{\left(\frac{-t^2}{2}\right)}}{\pi^{1/4}}.
\end{align}
These recurrence relations make the Hermite functions easy to implement and allow the explicit calculation of the derivative operator when transforming a CT LTI system, as shown in Section \ref{Differentiation operator}.
\\Figure \ref{fig: Hermite functions up to order 10.} shows the first 10 Hermite basis functions.

\subsection{Differentiation operator}
\label{Differentiation operator}
Subsection \ref{Subsec: Construction of CT-DT transformations} shows how expanding signals by means of Schauder bases transform operators in such a way that they act in $\ell^2 \to \ell^2$. In particular, as concerns system (\ref{eq: CT LTI dynamics in innovation form with explicit derivative}), the derivative - or differential - operator is the one that in practice is affected by the Schauder basis expansions. Hermite basis functions allow one to explicitly calculate the transformed differential operator due to their recurrence relations (\ref{eq: recurrence relations for Hermite functions}). Indeed, for the $n$-th normalized Hermite function, the following relation can be drawn:
\begin{equation}
\label{derivative of hermite function}
\frac{d}{dt}h_n(t) = \frac{1}{\sqrt{2}} \left(\sqrt{n} h_{n-1}(t) - \sqrt{n+1} h_{n+1}(t)\right).
\end{equation}
This implies that the derivative operator connects the state $n$ to the states $n-1$ and $n+1$. 
Now, let a function $f(t): \mathbb{R} \to \mathbb{R}^{n_f}$ be written as a series expansion:
\begin{equation}
\label{eq: function expanded with hermite bases}
f(t) = \sum_{n=0}^\infty{f_n h_n(t)},
\end{equation}
where each coefficient $f_n \in \mathbb{R}^{n_f}$ is obtained as follows:
\begin{equation}
f_n = \langle h_n(t), f(t) \rangle = \int_{\mathbb{T}} f(t) h_n(t) \, dt.
\end{equation}
Differentiating function (\ref{eq: function expanded with hermite bases}) allows one to extend the derivative operator to the discrete domain. First, it is useful to define the infinite-dimensional Hermite basis stack $\boldsymbol{h}(t): \mathbb{R} \to \mathbb{R}^{\infty}$ as:
\begin{equation}
\label{eq: function of hermite bases}
\boldsymbol{h}(t) \doteq \begin{bmatrix}
h_0(t) \\
h_1(t) \\
\vdots \\
\end{bmatrix}.
\end{equation}
By inserting equation (\ref{eq: function expanded with hermite bases}) into equation (\ref{derivative of hermite function}), and using (\ref{eq: function of hermite bases}), one gets:
\begin{equation}
\label{eq: time derivative with Hermite expansions}
\begin{split}
\frac{d}{dt} f(t) &= \sum_{n=0}^\infty{f_n \left(\frac{d}{dt}h_n(t)\right)} \\
&= \sum_{n=0}^\infty{f_n \left( \frac{1}{\sqrt{2}} \left(\sqrt{n} h_{n-1}(t) - \sqrt{n+1} h_{n+1}(t)\right)\right)} \\
&= \underbar{f}\mathcal{D}\boldsymbol{h}(t),
\end{split}
\end{equation}
where $\underbar{f}$ contains the projection coefficients and is defined as:
\begin{equation}
    \label{eq: hermite coefficients}
    \underbar{f} = \begin{pmatrix} \underbar{f}_{1} \\ \underbar{f}_{2}\\ \vdots \\ \underbar{f}_{n_f} \end{pmatrix} = \begin{pmatrix} f_{1,0} & f_{1,1} & \cdots \\ f_{2,0} & f_{2,1} & \cdots \\ \vdots & \vdots & \vdots \\ f_{n_f,0} & f_{n_f,1} & \cdots \end{pmatrix},
\end{equation}
each row corresponding to the coefficients of the $i$-th output of $f(t)$, while $\mathcal{D}$ is the derivative operator, which can be written explicitly in matrix form as follows:
\begin{equation}
\mathcal{D} = \begin{bmatrix} 0 & -\sqrt{1/2} & 0 & 0 & \cdots \\ \sqrt{1/2} & 0 & -\sqrt{2/2} & 0 &  \cdots \\ 0 & \sqrt{2/2} & 0 & -\sqrt{3/2} & \cdots  \\ 0 & 0 & \sqrt{3/2} & 0 & \ddots \\ \vdots & \vdots & \vdots & \ddots & \ddots \end{bmatrix}.
\end{equation}
One can easily verify that $\mathcal{D} = \tilde{\Phi}^\top$, where $\tilde{\Phi}$ is the one in (\ref{eq: transformed general operator}), with $\Phi = d(\cdot)/dt$ and $\{b_i\}_{i \in \mathbb{N}} = \{h_i\}_{i \in \mathbb{N}}$. The transpose is due to the coefficients in Subsection \ref{Subsec: Construction of CT-DT transformations} being stack as vectors. 
It is clear how to exploit relation (\ref{eq: time derivative with Hermite expansions}) if one considers the system in innovation form as in (\ref{eq: CT LTI dynamics in innovation form with explicit derivative}).
Indeed, by applying expansion (\ref{eq: function expanded with hermite bases}) to each time-dependent signal and relation (\ref{eq: time derivative with Hermite expansions}) to $\frac{d}{dt}\hat{x}(t)$, one obtains the following:
\begin{equation}
\label{eq: CT LTI dynamics with Hermite expansions and h(t)}
\begin{split}
\underbar{x} \mathcal{D} \boldsymbol{h}(t) = A \underbar{x} \boldsymbol{h}(t) + B \underbar{u} \boldsymbol{h}(t) + K\underbar{e}\boldsymbol{h}(t) \\
\underbar{y} \boldsymbol{h}(t) = C \underbar{x} \boldsymbol{h}(t) + D \underbar{u} \boldsymbol{h}(t) + \underbar{e}\boldsymbol{h}(t)
\end{split} \quad ,
\end{equation}
where $\underbar{x}$, $\underbar{u}$ and $\underbar{y}$ contain the projections on the Hermite bases of the corresponding signals. Thus, the transformed equations can be written as:
\begin{equation}
\label{eq: CT LTI dynamics with Hermite expansions}
\begin{split}
\underbar{x} \mathcal{D} = A \underbar{x} + B \underbar{u} + K\underbar{e}  \\
\underbar{y} = C \underbar{x} + D \underbar{u} + \underbar{e}
\end{split} \quad ,
\end{equation}
where innovation $\underbar{e}$ is a white noise signal.
It should be noted that, given the vector function $f(t)$ and its Hermite representation $\underbar{f}$ (equation (\ref{eq: hermite coefficients})), and given the sampling time instants $t_0, t_1, \hdots, t_N$ under which $f(t)$ is sampled, it is possible to reconstruct the values of the sampled signal $f(t)$ starting from its projections $\underbar{f}$ as follows: 
\begin{equation}
\label{eq: signal reconstruction}
\begin{pmatrix} \hat{f}(t_0) & \hat{f}(t_1) & \cdots & \hat{f}(t_N) \end{pmatrix} = \underbar{f} \underbar{H},
\end{equation}
where $\underbar{H}$ is the sampled Hermite basis matrix:
\begin{equation}
\underbar{H} = \begin{bmatrix} h_0(t_0) & h_0(t_1) & \hdots & h_0(t_N) \\ h_1(t_0) & h_1(t_1) & \hdots & h_1(t_N) \\ \vdots & \vdots & \vdots & \vdots & \end{bmatrix}. \label{DTCTTransform}
\end{equation}
In practice, a maximum order $n_{max}$ is considered for the Hermite expansion, with implications on the effective support described in Section \ref{Effective support and scaling}. Let $n_c = n_{max}+1$, accounting for the order 0 term. Thus:
\begin{equation}
\underbar{f} \in \mathbb{R}^{n_f\times n_c}, \mathcal{D} \in \mathbb{R}^{n_c\times n_c}, \underbar{H} \in \mathbb{R}^{n_c \times (N+1)}.
\end{equation}

\section{The HD-PBSID method}
\label{sec:hdpbsid}

\subsection{Effective support and scaling}
\label{Effective support and scaling}

\begin{figure}[htbp]
    \centering
    \includegraphics[width=\columnwidth]{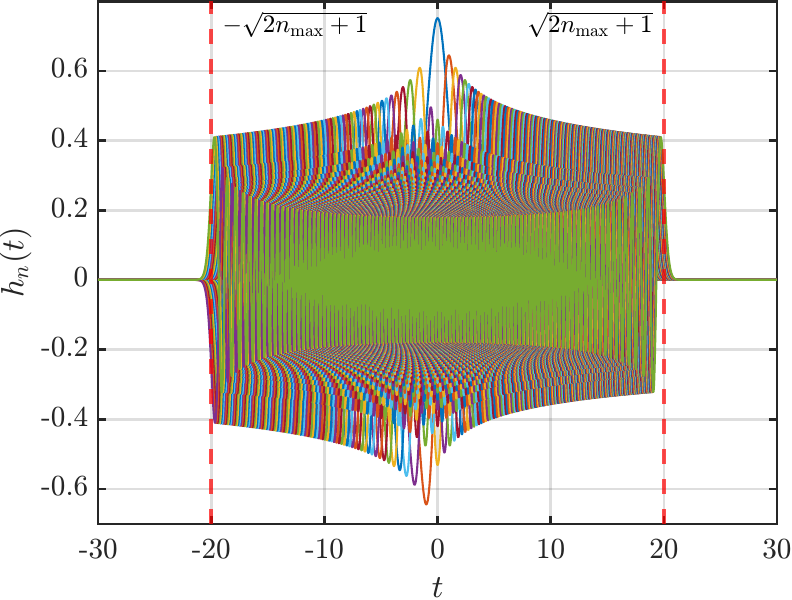}
    \caption{Hermite basis of order 200, with related numerically-significant region.}
    \label{fig:hermite-base}
\end{figure}

The proposed method employs a truncated Hermite expansion of the input and output signals.
Specifically, consider the basis $\mathcal{B} = \left\{h_n\right\}_0^{n_{max}}$, which contains $n_{max} + 1$ Hermite functions. 
While the full Hermite family forms a complete orthonormal basis of $\mathcal{L}^2(\mathbb{R})$, the truncated basis $\mathcal{B}$ provides only a finite-dimensional approximation. In practice, this approximation is most accurate over the region where the selected Hermite functions are numerically significant.  
Since - as a general property of Hermite functions - every Hermite function rapidly decays for $t$ being outside the domain $\left[-\sqrt{2n+1}, \sqrt{2n+1}\right]$, $n$ being the order of the $n$-th basis, it follows that the domain $\left[-\sqrt{2n_{max}+1}, \sqrt{2n_{max}+1}\right]$ represents the region - the so-called ``effective support"- outside of which all the $n_{max}$ functions assume negligible values.  
In Figure \ref{fig:hermite-base}, a basis of Hermite functions up to order 200 is displayed: thus, all functions tend to vanish outside the domain $\left[-\sqrt{401}, \sqrt{401}\right]$, which indeed constitutes the effective support. 

Consider an experiment taking place in the time interval $t \in \left[0, t^f\right]$, where $t^f = t^u_{N_u} \equiv t^y_{N_y}$. 
Then, input and output signals $u$ and $y$ should be time shifted so that their time vector is symmetric, that is, so that both signals are defined in a support $[-\tau, \tau]$, where $\tau = t^f/2$. Furthermore, it is convenient that any function $f(t)$ to be projected is contained within the effective support of $\mathcal{B}$. For that purpose, introduce the time change of variable 
\begin{equation}
t' = t \frac{\sqrt{2n_{max}+1}}{\tau} = \frac{t}{\alpha},
\end{equation}
so that $f(t) \rightarrow f(t')$. This transformed function fits the effective support of the considered basis. Note that this affects system (\ref{eq: CT LTI dynamics in innovation form with explicit derivative}), so that:
\begin{equation}
\frac{d}{dt} \hat{x} = A\hat{x} + Bu + Ke \rightarrow \frac{d}{dt'} \hat{x} = \alpha A \hat{x} + \alpha B u + \alpha Ke. \label{tScale}
\end{equation}
In order to obtain (\ref{eq: CT LTI dynamics with Hermite expansions}) for the Hermite subspace $\mathcal{B}$, the input and output signals are scaled and projected into $n_{max} + 1$ Hermite basis functions. The Hermite domain projection of the $m$-th dimension of each signal is thus given by
\begin{align}
\underbar{u}_{m n} & = \int_{-\sqrt{2n_{max}+1}}^{\sqrt{2n_{max}+1}}{u_m\left(t'\right) h_n(t') dt'} \\
\underbar{y}_{m n} & = \int_{-\sqrt{2n_{max}+1}}^{\sqrt{2n_{max}+1}}{y_m\left(t'\right) h_n(t') dt'}
\end{align}
for $n = 0, 1, \hdots, n_{max}$. 

\subsection{Modified differentiation operator}

Consider the transformed system in innovation form (\ref{eq: CT LTI dynamics with Hermite expansions}). By writing the innovation as a function of the input and of the output, one obtains:
\begin{equation}
\underbar{x} \mathcal{D} = \left(A-KC\right) \underbar{x} + \begin{bmatrix} B - KD & K \end{bmatrix} \begin{bmatrix} \underbar{u} \\ \underbar{y} \end{bmatrix} = \bar{A} \underbar{x} + \bar{B} \underbar{z},
\end{equation}
where $\bar{A} = A-KC$, $\bar{B} =  [B - KD \quad K]$ and $\underbar{z} = [ \underbar{u}^\top \quad \underbar{y}^\top]^\top$.
Note that $\bar{A}$ is stable in the CT sense, that is, it has only eigenvalues with negative real part. 
By accounting for the time scaling described in \eqref{tScale}, one obtains the following:
\begin{equation}
\label{eq: Hermite state equation with alpha}
\underbar{x} \mathcal{D} = \alpha \bar{A} \underbar{x} + \alpha \bar{B} \underbar{z}.
\end{equation}
This equation is not suitable for direct application of the PBSID algorithm, as it requires the eigenvalues of $\alpha \bar{A}$ to be within the unit circle, which is not granted.
This can be effectively handled by introducing a modified differentiation operator, namely,
\begin{equation}
\mathcal{D}' = \left(\frac{\mathcal{D}}{\alpha} + \beta I\right) \frac{1}{\gamma},
\end{equation}
for $\beta \ge 0$ and $ \gamma > 0$. The matrix components are thus modified as follows:
\begin{equation}
\mathcal{D}' = \begin{bmatrix} \frac{\beta}{\gamma} & \frac{-\sqrt{1/2}}{ \alpha\gamma} & 0 & 0 & \cdots \\ \frac{\sqrt{1/2}}{\alpha \gamma} & \frac{\beta}{\gamma} & \frac{-1}{\alpha \gamma} & 0 &  \cdots \\ 0 & \frac{1}{\alpha \gamma} & \frac{\beta}{\gamma} & \frac{-\sqrt{3/2}}{\alpha \gamma} & \cdots  \\ 0 & 0 & \frac{\sqrt{3/2}}{\alpha \gamma} & \frac{\beta}{\gamma} & \ddots \\ \vdots & \vdots & \vdots & \ddots & \ddots \end{bmatrix},
\end{equation}
with dimension $(n_{max}+1)\times(n_{max}+1)$.
Thus, equation (\ref{eq: Hermite state equation with alpha}) can be rewritten as follows:
\begin{equation}
\label{eq: LTI dynamics modified for stability}
\underbar{x} \mathcal{D}' = \frac{\bar{A} + \beta I}{\gamma} \underbar{x} + \frac{\bar{B}}{\gamma} \underbar{z} = \bar{A}' \underbar{x} + \bar{B}' \underbar{z}.
\end{equation}

\begin{figure}[htbp]
    \centering
    \includegraphics[width=\columnwidth]{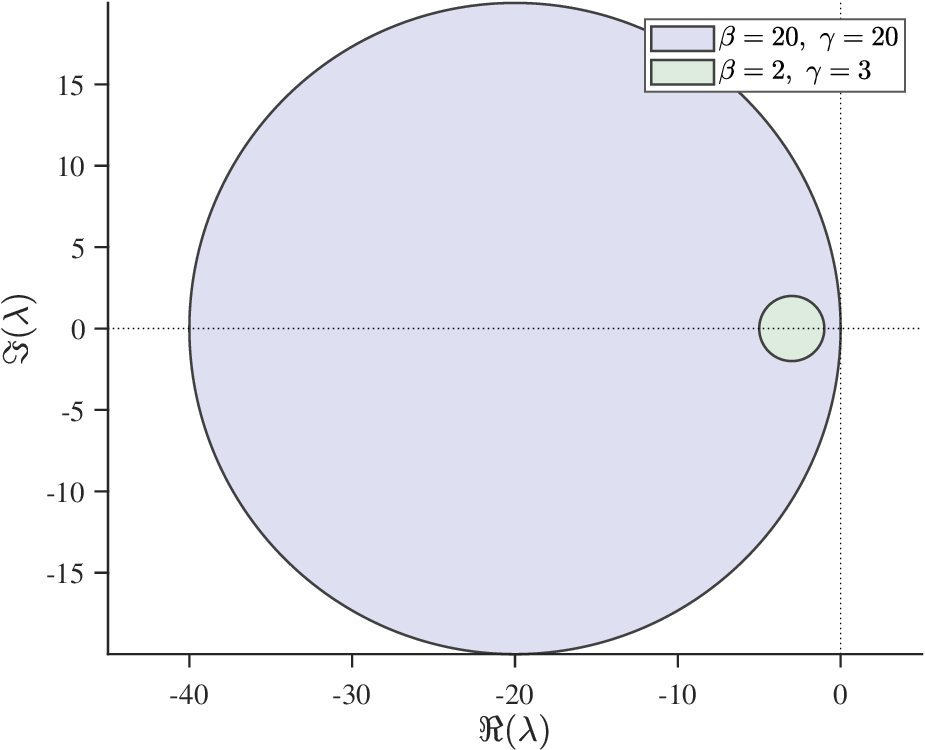}
    \caption{Stability region in the complex plane for the modified differentiation operator.}
    \label{fig:stab-region}
\end{figure}

If the eigenvalues of $\bar{A}$ are contained in a ball $B_{\beta,\gamma}$ with center $-\beta$ on the real axis and radius $\gamma$, as depicted in Figure \ref{fig:stab-region}, then the transformation from $\bar{A}$ to $\bar{A}'$ implicit in equation \eqref{eq: LTI dynamics modified for stability} guarantees that the eigenvalues of $\bar{A}'$ lie inside the unit circle.

\subsection{Integrator form and PBSID algorithm}

The PBSID algorithm can be applied by first differentiating \eqref{eq: LTI dynamics modified for stability} multiple times, leading to:
\begin{equation}
\underbar{x} \mathcal{D}'^p = \bar{A}'^p \underbar{x} + \bar{B}' \underbar{z} \mathcal{D}'^{p-1} + \bar{A}' \bar{B}' \underbar{z} \mathcal{D}'^{p-2}  + \cdots +  \bar{A}'^{p-1} \bar{B}' \underbar{z},
\end{equation}
where $p$ is the so-called past window length.
By exploiting $\bar{A}'^p \approx 0$, the following equation is obtained:
\begin{equation}
\underbar{x} \mathcal{D}'^p \simeq \begin{bmatrix} \bar{A}'^{p-1} \bar{B}' & \cdots &\bar{A}' \bar{B}' & \bar{B}' \end{bmatrix} \begin{bmatrix} \underbar{z} \\ \vdots \\ \underbar{z}\mathcal{D}'^{p-2} \\ \underbar{z}\mathcal{D}'^{p-1} \end{bmatrix},
\label{eq:statediff}
\end{equation}
where $ \mathcal{K}^p = \begin{bmatrix} \bar{A}'^{p-1} \bar{B}' & \cdots &\bar{A}' \bar{B}' & \bar{B}' \end{bmatrix}$ is the reversed extended controllability matrix.
Multiplying equation \eqref{eq:statediff} by $\mathcal{D}'^{-p}$ (\ie, integrating it $p$ times in the sense of the integral-like operator $\mathcal{D}'^{-1}$), the following expression is obtained,
\begin{equation}
\label{eq: integral version of state equation}
\underbar{x} \simeq \begin{bmatrix} \bar{A}'^{p-1} \bar{B}' & \cdots &\bar{A}' \bar{B}' & \bar{B}' \end{bmatrix} \begin{bmatrix} \underbar{z} \mathcal{D}'^{-p} \\ \vdots \\ \underbar{z} \mathcal{D}'^{-2}  \\ \underbar{z} \mathcal{D}'^{-1} \end{bmatrix}.
\end{equation}
Clearly, each $m$-th term $\underbar{z} \mathcal{D}'^{-m}$ is a filtered quantity, which does lead to numerical issues or noise amplification.
Letting now: 
\begin{equation}
\mathcal{Z} = \begin{bmatrix} \underbar{z} \mathcal{D}'^{-p} \\ \vdots \\ \underbar{z} \mathcal{D}'^{-2}  \\ \underbar{z} \mathcal{D}'^{-1} \end{bmatrix},
\end{equation}
it is possible to concisely rewrite the state sequence (\ref{eq: integral version of state equation}) as follows:
\begin{equation}
\label{eq: concise integral version of state equation}
\underbar{x} \simeq \mathcal{K}^p\mathcal{Z}.
\end{equation}
The output equation can be in turn written as:
\begin{equation}
\underbar{y} = C \mathcal{K}^p\mathcal{Z} + D  \underbar{u} + \underbar{e}.
\end{equation}
In order to estimate $C \mathcal{K}^p$ and $D$, the following least-squares problem has to be solved:
\begin{equation}
\label{eq:first least squares}
\min_{C \mathcal{K}^p, D} \left\| \underbar{y} - C \mathcal{K}^p \mathcal{Z} - D\underbar{u} \right\|_F
\end{equation}
where $\left\| \cdots \right\|_F$ represents the Frobenius norm.
It is now useful to define the following extended observability matrix:
\begin{equation}
\label{eq:extended observability matrix}
\Gamma^f = \begin{bmatrix} C \\ C\bar{A}' \\ \vdots \\ C \bar{A}'^{f-1} \end{bmatrix},
\end{equation}
where $f$ is the so-called future window length.
The product of $\Gamma^f$ and $\mathcal{K}^p$ leads to the following triangular upper block matrix:
\begin{equation}
\label{eq:simplified observability times controllability}
\Gamma^f \mathcal{K}^p \simeq 
\begin{bmatrix} 
C \bar{A}'^{p-1}\bar{B} & C \bar{A}'^{p-2}\bar{B}' & \cdots & C \bar{B}' \\
0 & C \bar{A}'^{p-1}\bar{B}' & \cdots & C \bar{A}' \bar{B}' \\
\vdots & & \ddots & \vdots \\
0 & \cdots & & C \bar{A}'^{p-1}\bar{B}'
\end{bmatrix},
\end{equation}
where $\bar{A}'^k \approx 0$ for $k \geq p$ is exploited.
The elements of this matrix can be computed from those of $\widehat{C \mathcal{K}^p}$, which is the estimate of $C \mathcal{K}^p$, obtained by solving the least-squares problem (\ref{eq:first least squares}).
By multiplying $\Gamma^f$ to both sides of Equation (\ref{eq: concise integral version of state equation}), one gets:
\begin{equation}
\label{eq:state sequence estimation}
\Gamma^f\underbar{x} \simeq \Gamma^f\mathcal{K}^p \mathcal{Z}, 
\end{equation}
where the right-hand side is known since $\widehat{\Gamma^f \mathcal{K}^p}$, the estimate of (\ref{eq:simplified observability times controllability}), is available. An estimate of the state sequence $\hat{\underbar{x}}$ can be computed up to a similarity transformation by applying the Singular Value Decomposition (SVD) to the right-hand side of (\ref{eq:state sequence estimation}):
\begin{equation}
\label{eq:SVD}
\Gamma^f\mathcal{K}^p \mathcal{Z} = \begin{bmatrix}
\mathcal{U}_n & \mathcal{U}_0 \end{bmatrix} \begin{bmatrix}
\Sigma_n & 0 \\ 0 & \Sigma_0 \end{bmatrix}
\begin{bmatrix}
\mathcal{V}^\top_n \\
\mathcal{V}^\top_0
\end{bmatrix},
\end{equation}
$\Sigma_n$ being the diagonal matrix containing the $n$ largest
singular values and $\mathcal{V}^\top_n$ the corresponding row space. An estimate of the order of the system can be provided by the number of singular values in $\Sigma_n$. The estimate of the state sequence reads as follows:
\begin{equation}
\label{eq:state sequence estimate via SVD}
\hat{\underbar{x}} \simeq \Sigma^{\frac{1}{2}}_n\mathcal{V}^\top_n = \Sigma^{-\frac{1}{2}}_n\mathcal{U}^\top_n\Gamma^f\mathcal{K}^p \mathcal{Z},
\end{equation}
which allows computing $C$ from:
\begin{equation}
\label{eq:second least squares}
\min_{C} \left\|  \underbar{y} - C \hat{\underbar{x}} - \hat{D} \underbar{u} \right\|_F .
\end{equation}
Thus, the estimate of the innovation sequence $\underbar{e}$ can be obtained:
\begin{equation}
\label{eq:innovation data matrix estimation}
\hat{\underbar{e}} = \underbar{y} - \hat{C} \hat{\underbar{x}} - \hat{D} \underbar{u} .
\end{equation}
The final step aims at computing the state space matrices and boils down to solving the following least-squares problem, from equation (\ref{eq: LTI dynamics modified for stability}):
\begin{equation}
\label{eq:third least squares}
\min_{A,B,K} \left\| \gamma \hat{\underbar{x}} \mathcal{D}' - A \hat{\underbar{x}} -\beta I \hat{\underbar{x}}  - B \hat{\underbar{u}} - K \hat{\underbar{e}} \right\|_F .
\end{equation}

\subsection{Parameters choice for the HD-PBSID method}
\label{sec: parameters of the algorithm}
The above-outlined HD-PBSID algorithm requires proper selection of $n_{max}$, $\gamma$ and $\beta$. 
\\Given $u(t)$ and $y(t)$, and their Hermite projections $\underbar{u}$ and $\underbar{y}$, the corresponding reconstructed signals $\hat{u}(t)$ and $\hat{y}(t)$ are obtained by applying equation (\ref{eq: signal reconstruction}), and thus they depend on $n_{max}$. Indeed, the reconstructed input (equivalently, the output) signal turns out to be:
\begin{equation}
\label{reconstructed input}
\hat{u}(t) = \sum_{n = 0}^{n_{max}}{h_n(t) \langle h_n(t), u(t) \rangle}.
\end{equation}
After undoing the time shift and the time scaling, $\hat{u}(t)$ and $\hat{y}(t)$ are assessed how closely they match $u(t)$ and $y(t)$, and how well they agree with their frequency spectra. 
To ensure a good reconstruction, and in consequence a successful identification, $n_{max}$ has to be properly tuned, for example by minimizing the errors between $u(t)$ and $\hat{u}(t)$ and between $y(t)$ and $\hat{y}(t)$.
\\A suitable choice of parameters $\gamma$ and $\beta$ is also crucial to ensure effective application of the PBSID algorithm. The eigenvalues $\lambda$ that are close to the boundary of $B_{\beta, \gamma}$ lead to an increased bias in the solution with respect to those near the center of the stability region. For this reason, it is convenient to use enlarged stability regions to improve the estimation and reduce bias.

\begin{rmk} 
\label{rmk1:convolutions}
The projection with Hermite functions leads to a matrix-like derivative operator, which turns out to be convenient for PBSID identification. Previous works on system identification \cite{Bergamasco2011} also exploit convolutions for obtaining discrete-time elements starting from continuous-time signals. 
In the convolution method, given a function $f(t) \in \mathcal{L}^2(\mathbb{R})$, the following operation is performed:
\begin{equation}
\left[h_n(t) * f(t) \right] (t) = \int_{-\infty}^{\infty}{h_n(t-\tau) f(\tau) d\tau }.
\end{equation}
Considering the symmetry property $h_n(-t) = (-1)^n h_n(t)$ for Hermite functions, the following can be written,
\begin{equation}
\begin{split}
&\left[h_n(t) * f(t) \right] (t_i) = (-1)^n \int_{-\infty}^{\infty}{h_n(\tau - t_i) f(\tau) d\tau } =\\&= (-1)^n \langle h_n(t-t_i), f(t) \rangle.
\end{split}
\end{equation}
This shows that convolutions are projections with alternating sign, leading to an equivalent coefficient expansion. This is consistent with the uniqueness of coefficients for an expansion of a given basis: convolutions lead to a set of coefficients only if they coincide with the ones provided by projections. Hermite functions thus provide a framework in which projections or convolutions can be equally used to obtain the coefficients of a series expansion of functions in $\mathcal{L}^2(\mathbb{R})$ over $\ell^2(\mathbb{N})$.
\end{rmk}

\section{Simulation results}
\label{sec:sims}

\subsection{Study of robustness and of parameters selection}
\label{subsec:secondorder}
\begin{figure}[htbp]
    \centering
    \begin{subfigure}[b]{\columnwidth}
    \includegraphics[width=\columnwidth]{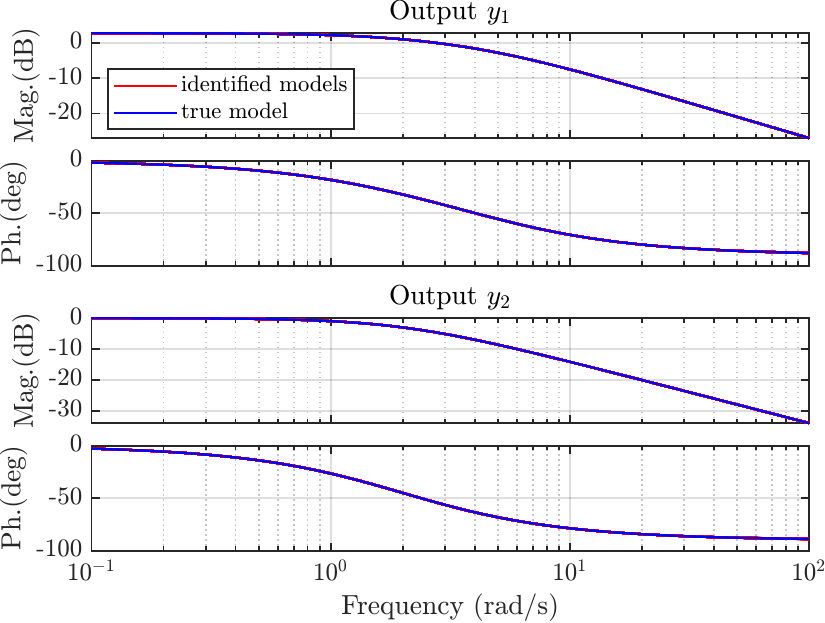}
    \caption{}
    \label{fig:sim_1}
    \end{subfigure}
    \begin{subfigure}[b]{\columnwidth}
    \centering
    \includegraphics[width=\columnwidth]{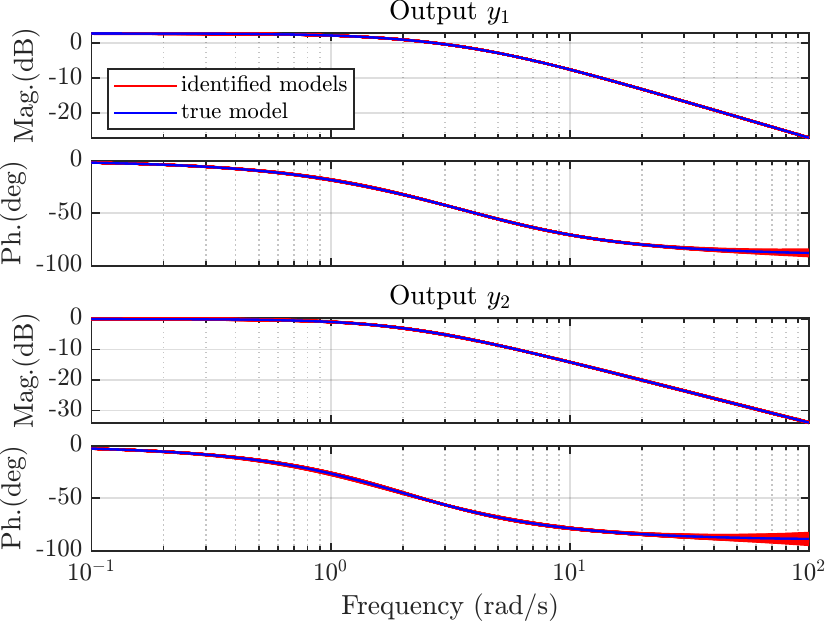}
    \caption{}
    \label{fig:sim_2}
    \end{subfigure}
   \caption{Bode diagram comparison between the real system and the HD-PBSID–identified systems obtained via a Monte Carlo simulation for (a) $\mathrm{SNR}=50$ dB and for (b) $\mathrm{SNR}=20$  dB.}
\end{figure}

\begin{figure}[htbp]
    \centering
    \begin{subfigure}[b]{\columnwidth}
    \includegraphics[width=\columnwidth]{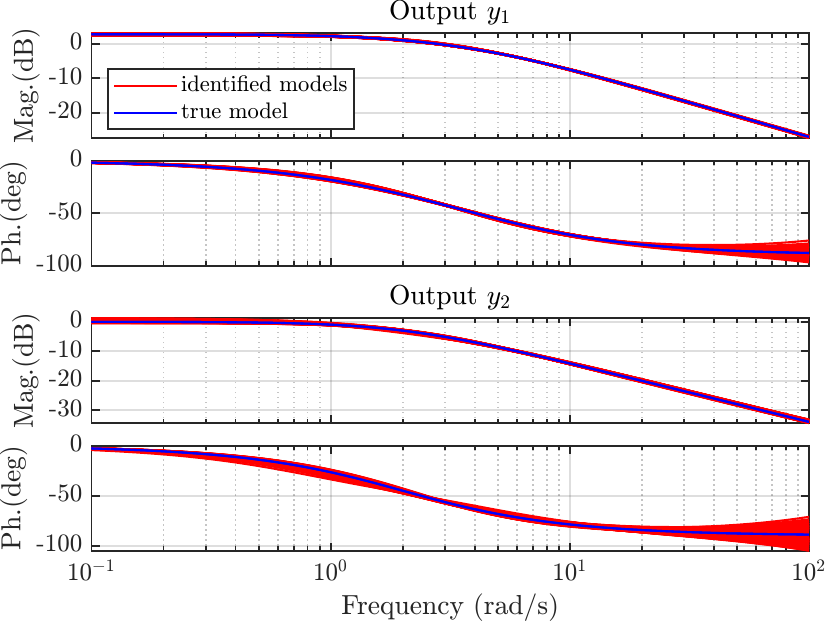}
    \caption{}
    \label{fig:sim_3}
    \end{subfigure}
    \begin{subfigure}[b]{\columnwidth}
    \centering
    \includegraphics[width=\columnwidth]{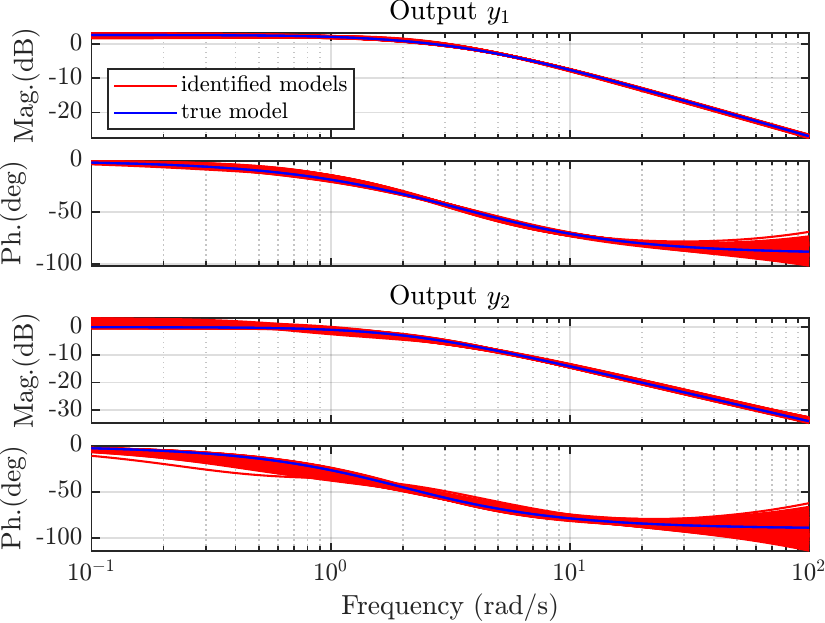}
    \caption{}
    \label{fig:sim_4}
    \end{subfigure}
   \caption{Bode diagram comparison between the real system and the HD-PBSID–identified systems obtained via a Monte Carlo simulation for (a) $\mathrm{SNR}=10$ dB and for (b) $\mathrm{SNR}=6$ dB.}
\end{figure}

\begin{figure}[htbp]
    \centering
    \begin{subfigure}[b]{\columnwidth}
    \includegraphics[width=\columnwidth]{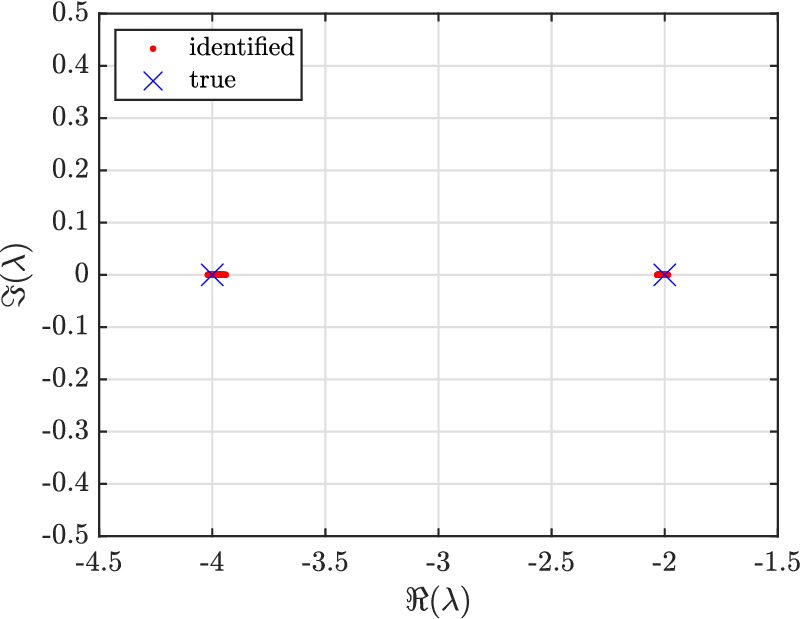}
    \caption{}
    \label{fig:sim_5}
    \end{subfigure}
    \begin{subfigure}[b]{\columnwidth}
    \centering
    \includegraphics[width=\columnwidth]{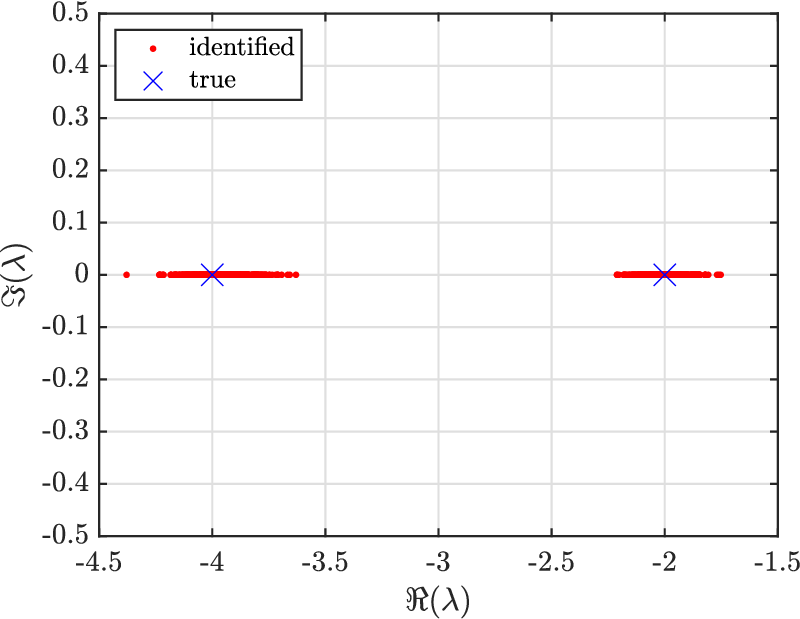}
    \caption{}
    \label{fig:sim_6}
    \end{subfigure}
   \caption{Eigenvalues comparison between the real system and the HD-PBSID–identified systems obtained via a Monte Carlo simulation for (a) $\mathrm{SNR}=50$ dB and for (b) $\mathrm{SNR}=20$ dB.}
\end{figure}

\begin{figure}[htbp]
    \centering
    \begin{subfigure}[b]{\columnwidth}
    \includegraphics[width=\columnwidth]{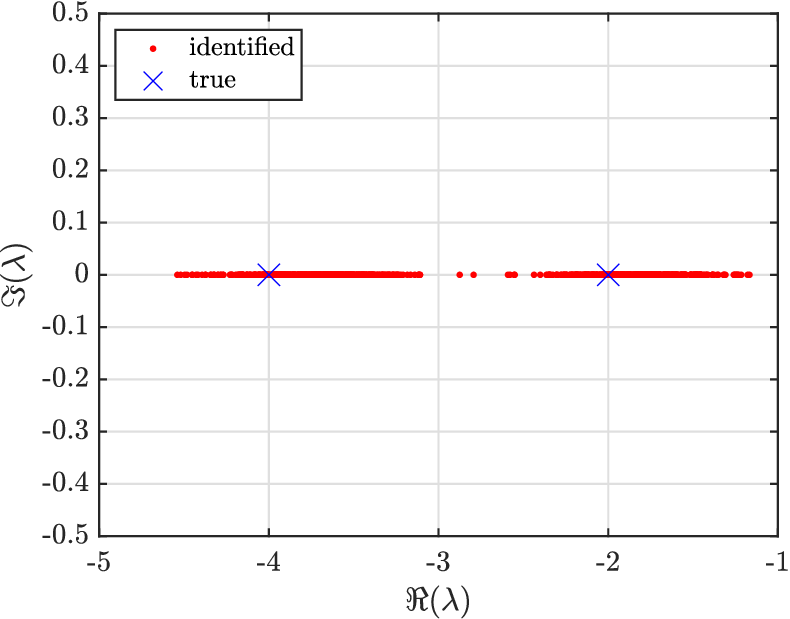}
    \caption{}
    \label{fig:sim_7}
    \end{subfigure}
    \begin{subfigure}[b]{\columnwidth}
    \centering
    \includegraphics[width=\columnwidth]{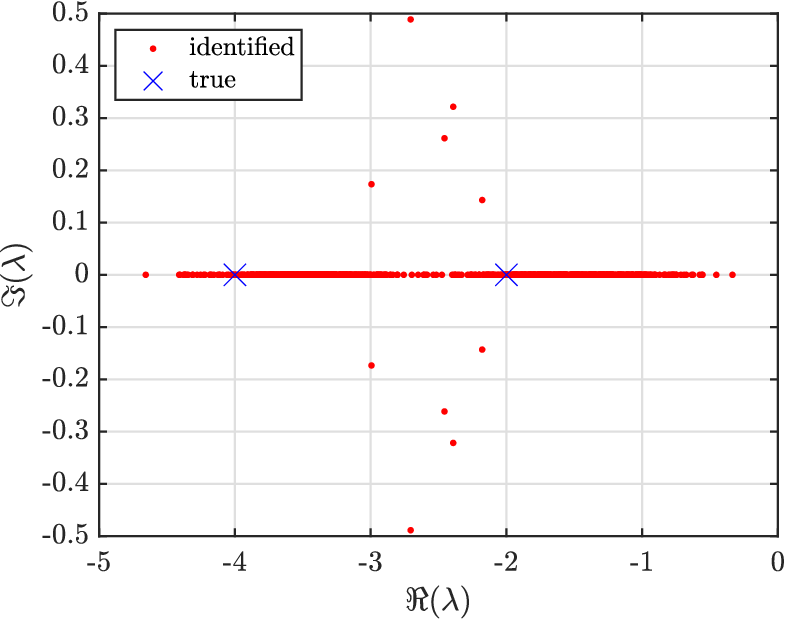}
    \caption{}
    \label{fig:sim_8}
    \end{subfigure}
   \caption{Eigenvalues comparison between the real system and the HD-PBSID–identified systems obtained via a Monte Carlo simulation for (a) $\mathrm{SNR}=10$ dB and for (b) $\mathrm{SNR}=6$ dB.}
\end{figure}

\begin{figure}[htbp]
    \centering
    \includegraphics[width=\columnwidth]{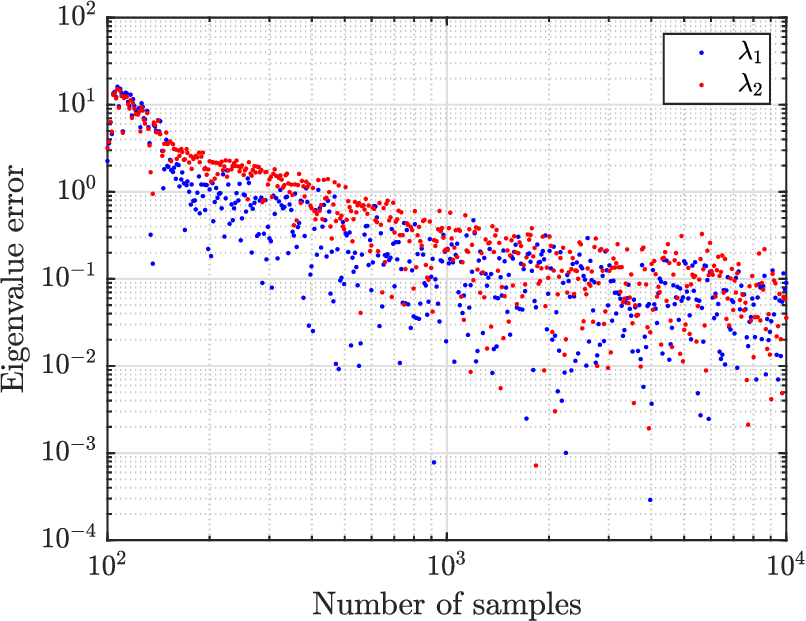}
    \caption{Error of the eigenvalues identified by the HD-PBSID for a varying number of samples.}
    \label{fig:sim_9}
\end{figure}

\begin{figure}[htbp]
    \centering
    \includegraphics[width=\columnwidth]{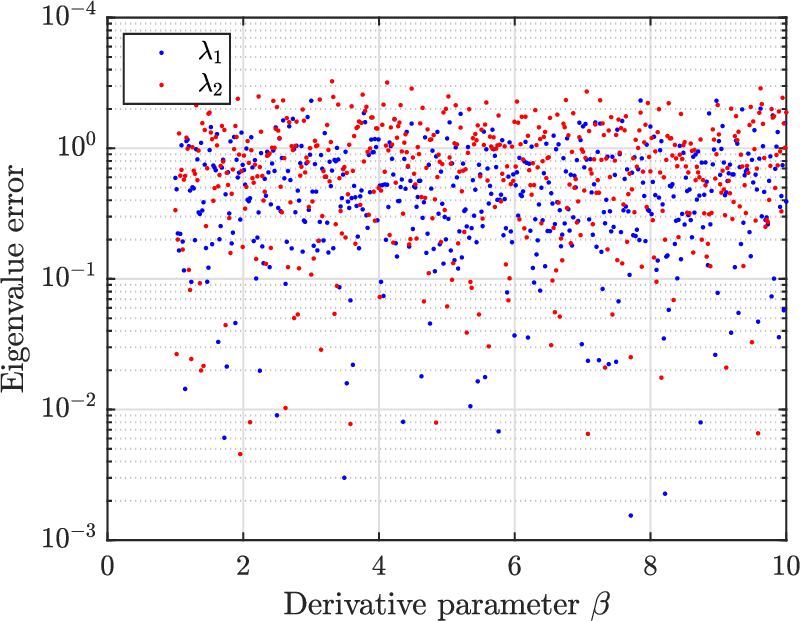}
    \caption{Error of the eigenvalues identified by the HD-PBSID for varying $\beta$ parameter.}
    \label{fig:sim_11}
\end{figure}

\begin{figure}[htbp]
    \centering
    \includegraphics[width=\columnwidth]{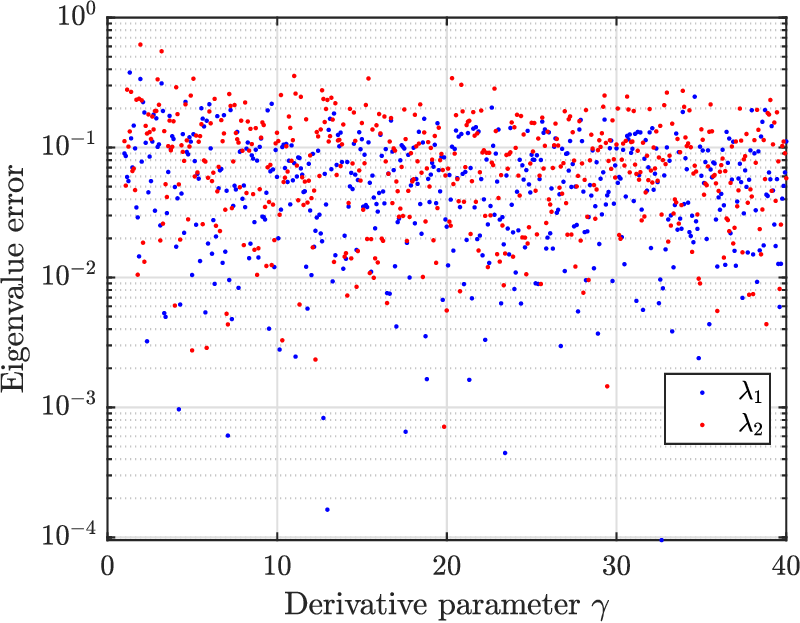}
   \caption{Error of the eigenvalues identified by the HD-PBSID for varying $\gamma$ parameter.}
   \label{fig:sim_12}
\end{figure}

In order to verify and validate computationally
the proposed identification algorithm,
the HD-PBSID method is tested against data generated by a  second-order system, given by:
\begin{equation*}
A = \begin{bmatrix} -2 & 1 \\ 0 & -4 \end{bmatrix},
B = \begin{bmatrix} 0.5 \\ 1 \end{bmatrix},
C = \begin{bmatrix} 1 & 4 \\ 2 & 1 \end{bmatrix},
D = \begin{bmatrix} 0  \\ 0 \end{bmatrix}.
\end{equation*}
A noise component of varying Signal to Noise Ratio (SNR) is introduced. The system is stable, with eigenvalues $\lambda_1 = -2$ and $\lambda_2 = -4$, and completely observable. The excitation is based on a linear sine sweep,
\begin{equation} 
\label{eq:linear sine sweep}
u(t) = \sin{\left[2\pi \left( f_1 t + \frac{f_2-f_1}{2T} t^2  \right)\right]}, 
\end{equation}
where $T = 10\mathrm{~s}$ is the duration of the experiment, while $f_1 = 0\mathrm{~Hz}$ and $f_2 = 8\mathrm{~Hz}$ are the minimum and maximum excitation frequencies, respectively. 
The maximum expansion order of the Hermite basis is $n_{max} = 250$.
The actual exciting function can be computed as in (\ref{reconstructed input}).
The differentiation coefficients are chosen as $\beta=3$ and $\gamma=20$.
Given this choice in terms of parameters and excitation signal, a Monte Carlo simulation of $500$ runs is performed by varying the noise randomness seed. The simulation is repeated for SNR = $\{6\, ,\, 10\,, \,20\,,\, 50\}$ dB.   
In Figures \ref{fig:sim_1}--\ref{fig:sim_4}, the Bode diagrams of the identified systems, together with the real one, are displayed for varying SNR values. Similarly, the modes of the identified dynamics are compared to the exact ones in Figures \ref{fig:sim_5}--\ref{fig:sim_8}. For low noise, all simulations converge closely to the real system, with a very good fit for the Bode plots. As the SNR decreases, the identified systems deviate from the actual dynamics, particularly at the extremes of the frequency range.  Excluding a small set of simulations for SNR = $6$ dB, almost all identified eigenvalues are real, as expected.
\\The effect of different parameters on the HD-PBSID algorithm is also evaluated. In particular, four parameters are changed independently: the sampling frequency, the expansion order $n_{max}$, and the stabilizing parameters $\beta$ and $\gamma$. In all cases, moderate-high noise with SNR $=20$ dB is introduced. 
\\The effect of the number of samples on the error of the identified eigenvalues is first analyzed, considering values between 100 and 10000 samples, as shown in Figure \ref{fig:sim_9}. As expected, more samples provide a better identification of the dynamics. However, the quality of the identified system does not improve significantly for more than 2000 samples. 
The effect of the choice of $n_{max}$ is further analyzed, for expansions from order 10 to order 1000. In this case, an expansion order lower than 30 produces a very poor identification accuracy, while increasing the order past 50 improves the accuracy only slightly. As explained in Section \ref{sec: parameters of the algorithm}, at low order significant information is lost when projecting on the Hermite domain. When the order reaches the maximum order excited by the system, increasing the size of the expansion only provides terms of a negligible effect. As all functions, once projected into the Hermite domain, belong to $\ell^2$, their coefficients must eventually fall to zero, allowing one to always define a maximum order.
\begin{figure}[htbp]
    \centering
    \includegraphics[width=\columnwidth]{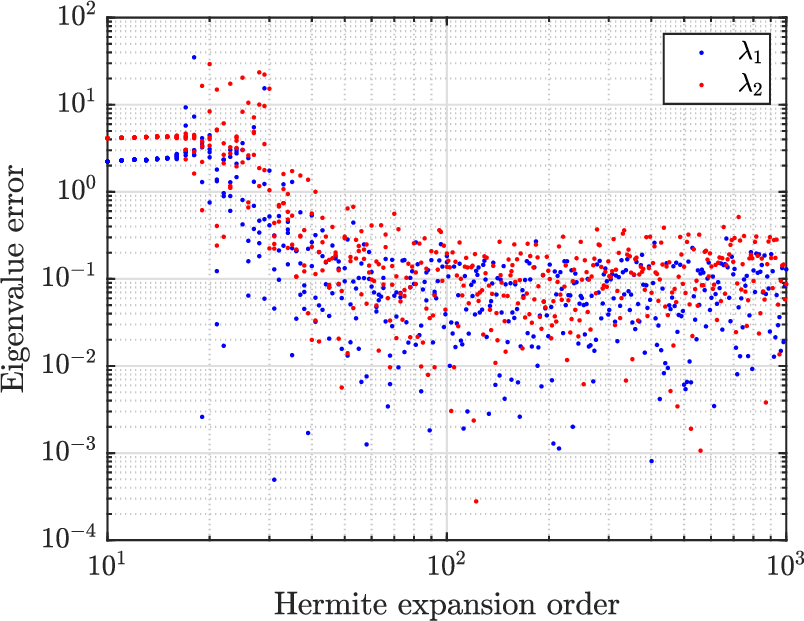}
    \label{fig:sim_10}
   \caption{Error of the eigenvalues identified by the HD-PBSID for a varying Hermite basis order.}
\end{figure}
Finally, the effect of $\beta$ and $\gamma$ is considered. As discussed in Section \ref{sec: parameters of the algorithm}, these parameters do not have a significant effect on the identification, as long as $\gamma$ is large enough and does not compromise numerical stability. Figure \ref{fig:sim_11} shows the error of the identified eigenvalues when $\beta$ is varied in the range $[1\,,\, 10]$, while $\gamma = 20$. It should be noted that $\lambda_1/\gamma = -2/20 = -0.1$ and $\lambda_2/\gamma = -4/20 = -0.2$ are smaller than one in magnitude, and the range of $\beta$ is chosen such that there is no combination $(\lambda + \beta)/\gamma$ outside the unit circle. In Figure \ref{fig:sim_12} $\gamma$ is varied in the range $[1\,,\, 40]$, while  $\beta = 3$. It should be pointed out that in this case both $(\lambda_1 + \beta)/\gamma$ and $(\lambda_2 + \beta)/\gamma$ are equal to $1/\gamma$ in absolute value. Hence, $(\lambda + \beta)/\gamma$ is within the unit circle as long as $\gamma > 1$. It follows that almost all simulations provide comparable errors, with a slight improvement when increasing $\gamma$, as this reduces the bias of the estimator by making $\bar{A}'^p$ converge faster.
\subsection{Comparison with the CT-PBSID algorithm}
\label{subsec:ctpbsidcomp}
\begin{figure}[htbp]
    \centering
    \begin{subfigure}[b]{\columnwidth}
    \includegraphics[width=\columnwidth]{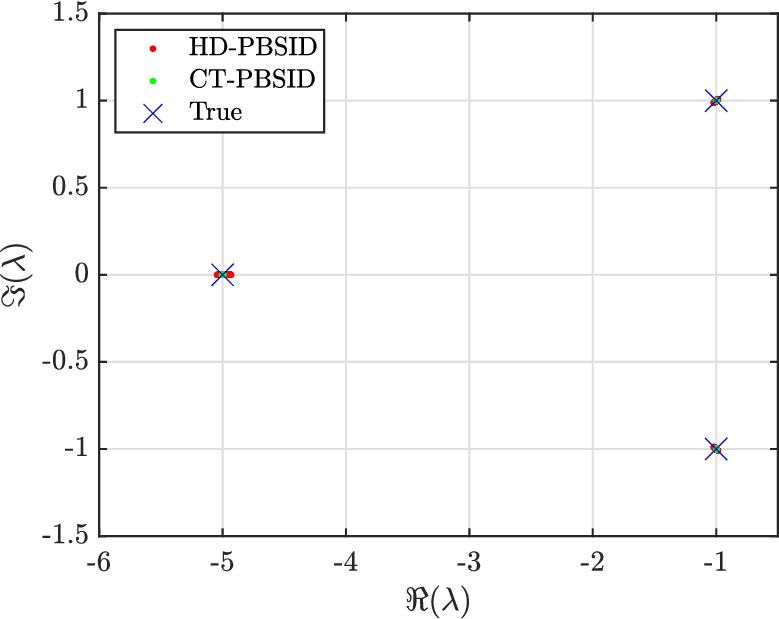}
    \caption{}
    \label{fig:comp_1}
    \end{subfigure}
    \begin{subfigure}[b]{\columnwidth}
    \centering
    \includegraphics[width=\columnwidth]{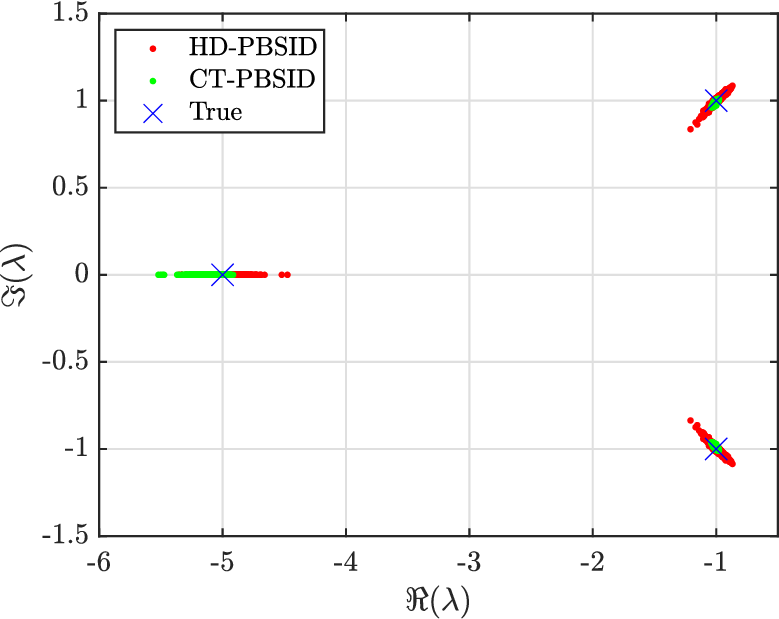}
    \caption{}
    \label{fig:comp_2}
    \end{subfigure}
   \caption{Eigenvalues comparison between CT-PBSID and HD-PBSID–identified systems obtained via a Monte Carlo simulation for (a) $\mathrm{SNR}=50$ dB and for (b) $\mathrm{SNR}=20$ dB.}
\end{figure}

\begin{figure}[htbp]
    \centering
    \begin{subfigure}[b]{\columnwidth}
    \includegraphics[width=\columnwidth]{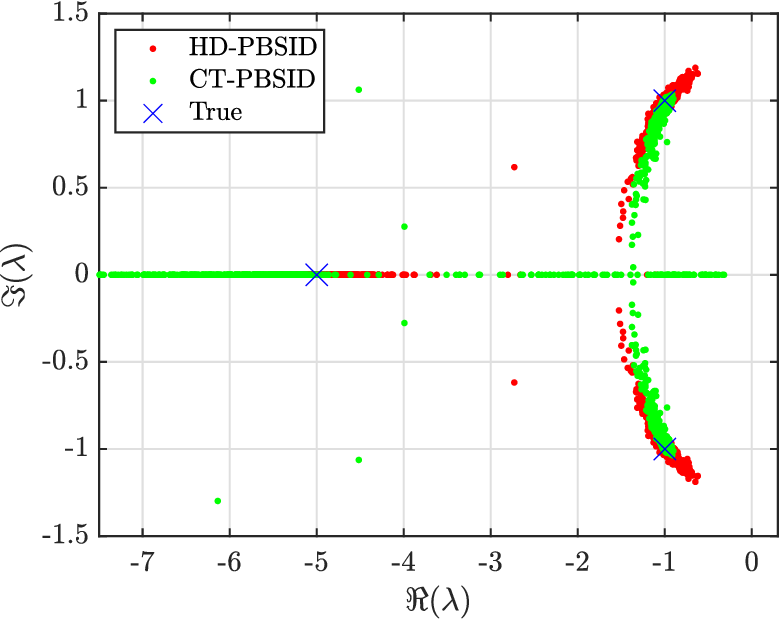}
    \caption{}
    \label{fig:comp_4}
    \end{subfigure}
    \begin{subfigure}[b]{\columnwidth}
    \centering
    \includegraphics[width=\columnwidth]{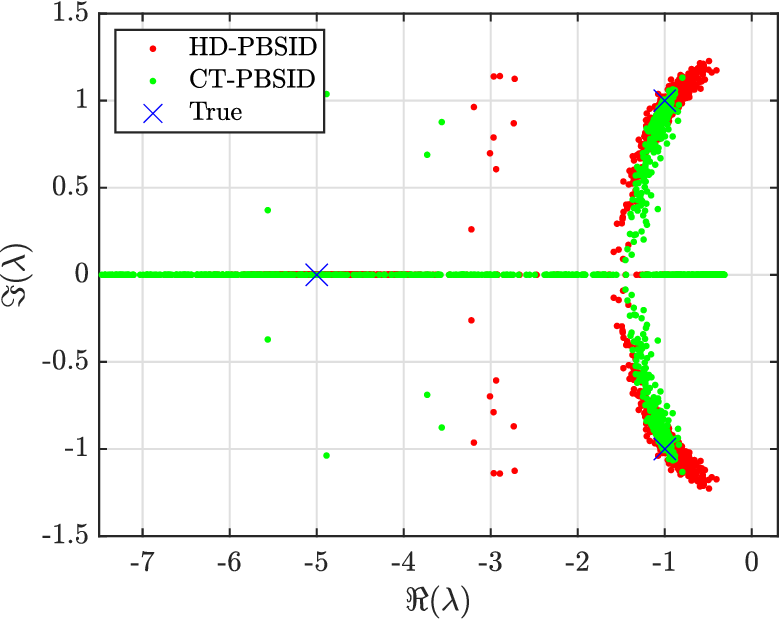}
    \caption{}
    \label{fig:comp_6}
    \end{subfigure}
   \caption{Eigenvalues comparison between the CT-PBSID and the HD-PBSID–identified systems obtained via a Monte Carlo simulation for (a) $\mathrm{SNR}=10$ dB and for (b) $\mathrm{SNR}=6$ dB.}
\end{figure}

In order to further validate the HD-PBSID method and assess its effectiveness, the algorithm is compared with the CT-PBSID method. 
In particular, the following system is considered for performing numerical simulations at different noise conditions and then using the input/output data for identifications with both the CT-PBSID and the HD-PBSID: 
\begin{align*}
A & = \begin{bmatrix} -0.2588 & 0.2115 & -9.5751 \\ 0.7629 &  -6.7412 & -10.4169 \\ 0 & 1 & 0 \end{bmatrix}, \;
B = \begin{bmatrix} -10.1647 \\ 450.71 \\ 0 \end{bmatrix}, \\
C & = \begin{bmatrix} 1 & 0 & 0 \\ 0 & 1 & 0 \\ 0 & 0 & 1 \\ -0.1068 & 0.1192 & 0 \end{bmatrix}, \;
D = \begin{bmatrix} 0 \\ 0 \\ 0 \\ -10.1647 \end{bmatrix}.
\end{align*}
The plant has the `true' eigenvalues $\lambda^t$ at $\lambda^t_{1} = -5$ and $\lambda^t_{2,3} = -1\pm i$. 
The excitation signal $u$ is designed by exploiting equation (\ref{reconstructed input}) as follows:
\begin{equation}
\label{input for comparison}
u(t) = \sum_{n = 0}^{\eta}{h_n(t) \langle h_n(t), u_S(t) \rangle},
\end{equation}
where $u_S(t)$ is the linear time sweep of equation (\ref{eq:linear sine sweep}), with $T = 10\mathrm{~s}$, $f_1 = 0\mathrm{~Hz}$ and $f_2 = 8\mathrm{~Hz}$, and $\eta = 250$. This is done to keep the discrepancy between $\hat{u}$ and $u$ small when $n_{max} \geq \eta$.
The parameters of HD-PBSID algorithm are set to $n_{max} = 300$, $\beta = 3$ and $\gamma = 30$. CT-PBSID with Laguerre projections depends on some parameters as well: the past and future window lengths, respectively set to $p=10$ and $f=10$, and the position of the Laguerre pole $a = 25$.  
A Monte Carlo simulation of $N_{MC} = 500$ trials is performed by varying the noise randomness seed. The simulation is repeated for SNR = $\{6\, ,\, 10\,, \,20\,,\, 50\}$ dB, with the results displayed in Figures \ref{fig:comp_1}--\ref{fig:comp_6}. 
Given the eigenvalues $\lambda^H$ estimated via HD-PBSID and $\lambda^L$ estimated via CT-PBSID, the bias and the standard deviation of the estimation error are computed and displayed in Table \ref{tab:snr_results}. For instance, considering the HD-PBSID, the bias of each $k$-th eigenvalue is computed as the absolute error of the mean:
\begin{equation}
\mathrm{Bias}^H_k
=
\left|
\frac{1}{N_{MC}}
\sum_{j=1}^{N_{MC}}
(\lambda^{H}_{k,j}
-
\lambda^{t}_k)
\right|
\end{equation}
The standard deviation (STD) is computed as follows:
\begin{equation}
\mathrm{STD}^{H}_k
=
\sqrt{
\frac{1}{N_{MC}-1}
\sum_{j=1}^{N_{MC}}
\left|
\lambda^H_{k,j}
-
\overline{\lambda}^H_{k}
\right|^2
}
\end{equation}
with $\overline{\lambda}^H_{k} = \frac{1}{N_{MC}}\sum_{j=1}^{N_{MC}}\lambda^H_{k,j}$. Analogous expressions are used for the CT-PBSID method.
\begin{table}[t]
\centering
\footnotesize
\caption{Bias and Standard Deviation (STD) comparisons.}
\label{tab:snr_results}
\begin{tabular}{cc|cc|cc}
\toprule
SNR & $\lambda$ 
& \multicolumn{2}{c}{Bias} 
& \multicolumn{2}{c}{STD} \\
\cmidrule(lr){3-4} \cmidrule(lr){5-6}
 &  & Hermite & Laguerre & Hermite & Laguerre \\
\midrule
\multirow{2}{*}{50}
 & $\lambda_1$   & 0.010883 & 0.000437 & 0.021611 & 0.001531 \\
 & $\lambda_{2,3}$ & 0.002407 & 5.949e-05 & 0.006298 & 0.000290 \\
\midrule
\multirow{2}{*}{20}
 & $\lambda_1$   & 0.024859 & 0.096078  & 0.1201 & 0.088157 \\
 & $\lambda_{2,3}$ & 0.006865 & 0.006385 & 0.061848 & 0.0116 \\
\midrule
\multirow{2}{*}{10}
 & $\lambda_1$   & 0.086345  & 2.4266   & 0.37473 & 7.6673 \\
 & $\lambda_{2,3}$ & 0.040319 & 0.20973  & 0.21217 & 0.34462 \\
\midrule
\multirow{2}{*}{6}
 & $\lambda_1$   & 0.15362  & 44.717    & 0.60203 & 271.48 \\
 & $\lambda_{2,3}$ & 1.8876   & 1.928   & 0.51834 & 1.8059 \\
\bottomrule
\end{tabular}
\end{table}
At low noise intensity (SNR $ = 50$ dB and SNR $ = 20$ dB), both algorithms provide estimates that are close to the true system eigenvalues, with small bias and STD. However, at higher noise intensity (SNR $ = 10$ dB and especially SNR $ = 6$ dB)  the CT-PBSID exhibits very large STDs and significant bias, indicating that only few identifications are reasonable. On the contrary, even at high noise intensities HD-PBSID provides accurate estimates, with small STD and bias, showing its robustness to different noise conditions.    
In general, the fact that the HD-PBSID originates from a rigorous non-redundant approach to projection of signals and operators, together with the considered input setup based on a lower order Hermite expansion, seems to improve the accuracy of the identification in very noisy scenarios. 
\section{Conclusions}
In this paper, the HD-PBSID method is presented. This new method is shown to allow direct identification of continuous-time matrices of LTI systems, without intermediate transformations. System identification performed on simulation data show the efficacy of the method and its robustness to changes both on the stabilizing parameters and on noise intensity. Improved standard deviation results, compared with the Laguerre-projection CT-PBSID method under increased noise intensity, indicate that HD-PBSID may be a promising technique for future industrial applications.   

\bibliographystyle{unsrt}
\bibliography{bibliography}          
                            
\end{document}